\documentclass[
preprint,
prd,tightenlines,showpacs,nofootinbib,
eqsecnum,
amsfonts,amsmath,amssymb]{revtex4-1}

\usepackage{bm}
\usepackage{graphicx} 
\usepackage{hyperref}
\usepackage{mathrsfs}

\newcommand{\ud}{\mathrm{d}}

\newcommand{\ph}[1]{\phantom{#1}}
\newcommand{\vecnab}{\bm{\nabla}}

\allowdisplaybreaks

\begin{document}

\title{Modified gravity approach based on a preferred time foliation}

\author{Luc \textsc{Blanchet}}\email{blanchet@iap.fr}
\affiliation{$\mathcal{G}\mathbb{R}\varepsilon{\mathbb{C}}\mathcal{O}$ Institut d'Astrophysique de Paris --- UMR 7095 du CNRS, \\ Universit\'e Pierre \& Marie Curie, 98\textsuperscript{bis} boulevard Arago, 75014 Paris, France}

\author{Sylvain \textsc{Marsat}}\email{marsat@iap.fr}
\affiliation{$\mathcal{G}\mathbb{R}\varepsilon{\mathbb{C}}\mathcal{O}$ Institut d'Astrophysique de Paris --- UMR 7095 du CNRS, \\ Universit\'e Pierre \& Marie Curie, 98\textsuperscript{bis} boulevard Arago, 75014 Paris, France}

\date{\today}

\begin{abstract}
We propose, in a heuristic way, a relativistic modified gravity model as an alternative to particle dark matter at galactic scales. The model is based on a postulated preferred time foliation described by a dynamical scalar field called the ``Khronon''. In coordinates adapted to the foliation it appears as a modification of general relativity violating local Lorentz invariance in a regime of weak gravitational fields. The model is a particular case of non-canonical Einstein-{\ae}ther theory, but in which the {\ae}ther vector field is hypersurface orthogonal. We show that this model recovers the phenomenology of the modified Newtonian dynamics (MOND) in the non-relativistic limit, and predicts the same gravitational lensing as general relativity but with a modified Poisson-type potential.
\end{abstract}

\pacs{95.35.+d,95.36.+x,04.50.Kd}

\maketitle

\section{Introduction}
\label{secI}

Modifying gravity in the regime of weak gravitational fields or accelerations, is the leitmotiv of MOND --- the Modified Newtonian Dynamics \cite{Milg1,Milg2,Milg3}. The MOND paradigm is very successful at solving the problem of the flat rotation curves of galaxies without the need of dark matter (see \cite{SandMcG02} for a review), and at explaining the correlations between the distribution and dynamics of ordinary matter versus dark matter \cite{TF,McG11}. However, it is based on a non-relativistic formula and needs to be generalized in order to be applied in cosmology and to the gravitational lensing observed at galaxy cluster scale \cite{GerbalDurret,Clowe,ASZF07}, which are part of the great successes of the standard cosmological paradigm $\Lambda$-CDM based on particle cold dark matter \cite{BHS05} and the cosmological constant $\Lambda$. 

We adopt for MOND the modification of the non-relativistic Poisson equation~\cite{BekM84},
\begin{equation}\label{MOND}
\vecnab \cdot \left[ \mu\left(\frac{g}{a_0}\right) \vecnab \phi
  \right] = 4\pi G \rho\;,  
\end{equation}
where $\rho$ is the density of ordinary (essentially baryonic) matter, $\phi$ is the gravitational potential, $\bm{g}=-\vecnab \phi$ is the gravitational field and $g = \vert\bm{g}\vert$ its ordinary Euclidean norm. The MOND function $\mu$ depends on the single argument $g/a_0$, where $a_0=1.2\times 10^{-10}\,\mathrm{m}/\mathrm{s}^2$ is a constant acceleration scale. In the limit of weak gravitational fields \textit{i.e.} $g\ll a_0$, the MOND function is linear, $\mu(g/a_0)=g/a_0$, while in the strong field regime $g\gg a_0$ (though non relativistic), $\mu$ tends to one so that \eqref{MOND} reduces to the usual Poisson equation.

Several relativistic extensions of MOND have been proposed (see \cite{BGef07,CFPS11} for reviews). The Tensor-Vector-Scalar (TeVeS) theory of Bekenstein and Sanders \cite{Sand97,Bek04,Sand05} extends general relativity with a time-like vector field and one scalar field. Einstein-{\ae}ther theories \cite{JM00,JM04} involve a unit time-like vector field which is non-minimally coupled to the metric, and were originally used to describe the phenomenology of local Lorentz invariance violation. Those theories, when modified to involve non-canonical kinetic terms, have been shown to provide interesting examples of relativistic MOND theories \cite{ZFS07,HZL08}. The vector field postulated in TeVeS is in fact analogous to the vector field in Einstein-{\ae}ther theories \cite{ZFS06,CWW08}. The cosmology in TeVeS and Einstein-{\ae}ther theories has been extensively studied \cite{SMFB06,LMB07,Sk08,ZZ10}. The parametrized post-Newtonian parameters, included the preferred frame parameters crucial in the presence of a vector field, have been worked out for both the canonical Einstein-{\ae}ther theory \cite{FJ06} and TeVeS \cite{Sagi09}. More recent proposals for relativistic MOND include a bimetric theory of gravity \cite{Milgbim1,Milgbim2}, a refinement of TeVeS using a Galileon ``$k$-mouflage'' to prevent deviations from general relativity at small distances \cite{BDEF11}, and a non-local extension of general relativity \cite{DEFW11}. One can also invoke some new exotic properties of the dark matter rather than modifying gravity \cite{BL08,BL09}.

On the other hand, much interest focused recently on the possible violation of the local Lorentz invariance in a completion of general relativity at high energy. This Ho\v{r}ava-Lifshitz approach \cite{Horava09} is motivated by the construction of a power-counting renormalizable quantum theory of gravity. The Lorentz invariance violation results from a preferred time foliation which is described in a $4$-dimensional covariant formulation by a dynamical scalar field sometimes called the ``Khronon'' \cite{BPS11}. A consistent extension of the (non-projectable version of the) Ho\v{r}ava gravity has been proposed \cite{BPS10,BPS11} to avoid some problems regarding the stability of the theory. In this extension the added terms crucially involve the acceleration of the congruence normal to the preferred family of hypersurfaces. It has been shown that the low energy limit of the extended Ho\v{r}ava gravity is equivalent to an Einstein-{\ae}ther theory in which the vector field is hypersurface-orthogonal \cite{Jacobson10,BPS11}. 

In the present paper, we revisit some works on Einstein-{\ae}ther theories with non-canonical kinetic terms (notably \cite{ZFS07,HZL08}), but in the framework of a preferred time foliation, inspired by these recent approaches to quantum gravity. We propose a relativistic MOND theory which modifies general relativity by terms depending on the acceleration in a way somewhat similar to the extended Ho\v{r}ava gravity \cite{Horava09,BPS10,BPS11}. However, our motivation is quite different from that in \cite{Horava09,BPS10,BPS11}, as it is purely classical, and we are not concerned with the problem of quantizing gravity. In particular, the violation of Lorentz invariance will be effective at low energy, for very weak gravitational fields. Our aim is to provide a phenomenologically viable alternative to the particle cold dark matter (at least at galactic scales). 

The modification we propose consists of introducing in the Lagrangian of general relativity a single free function of the norm $a$ of the acceleration of the congruence normal to the preferred hypersurfaces. (Unlike in \cite{BPS10,BPS11}, we are not considering a finite number of terms including this acceleration, each one with a coupling constant scaling with the Planck mass, but a fully non-linear function.) The theory will be a particular case of non-canonical Einstein-{\ae}ther theory but in which the {\ae}ther field is hypersurface orthogonal. We shall present the theory in both a 3+1 decomposition and $4$-dimensional covariant form, and shall explicitly prove the equivalence of the two formulations. In agreement with general studies on non-canonical Einstein-{\ae}ther theories \cite{ZFS07,HZL08}, we will show that it is possible, by specifying the asymptotic behaviour of the function of the acceleration when $a\ll a_0$, to recover the Bekenstein-Milgrom equation \eqref{MOND} in the non-relativistic limit. The theory will also be viable regarding the gravitational lensing of photons, which will be the same as in general relativity but with a Newtonian potential obeying the modified Poisson equation.\footnote{While this work was in preparation, a paper based on similar ideas appeared on the archives \cite{Sand11}. However, the field equations in that paper are in disagreement with ours.}

The plan of this paper is as follows. In Section \ref{secII} we present the $4$-dimensional version of the theory, introducing the dynamical Khronon field defining the space-time foliation, and investigating the resulting field equations. In Section \ref{secIII} we switch to an alternative 3+1 point of view, for which the time coordinate coincides with the Khronon field, and where the new terms associated with the acceleration in the action may be interpreted as Lorentz invariance breaking terms due to a preferred-frame effect. In Section \ref{secIV}, we investigate the non-relativistic limit and the conditions under which one can retrieve the non-relativistic MOND equation \eqref{MOND}. Section \ref{secV} contains some concluding remarks. Finally, Appendix \ref{appA} is devoted to the proof of equivalence between the $4d$ covariant formulation and the 3+1 formalism, while Appendix \ref{appB} presents the canonical decomposition of the energy-momentum tensor associated with the Khronon field when interpreted as a matter field. 

\section{Covariant formulation}
\label{secII}

In the 4-dimensional covariant formulation, we extend general relativity by means of the additional Khronon scalar field denoted by $\tau$. We define a preferred foliation of space-time according to the constant-$\tau$ hypersurfaces, so we will require $\tau$ to be smooth and free of extrema. In Section \ref{secIII} below, we consider a frame adapted to this foliation in which the time coordinate $t$ will coincide with $\tau$, but in the present Section we investigate the fully covariant formalism. 

As in the usual 3+1 formalism, we introduce the unit timelike vector $n^{\mu}$, oriented toward the future ($n^0>0$), orthogonal to the constant-$\tau$ hypersurfaces, and normalized to $g_{\mu\nu}n^\mu n^\nu=-1$.\footnote{We adopt the sign conventions of \cite{MTW}, and notably the $(-+++)$ signature for the metric. Greek indices are space-time indices, while Latin are space indices. We pose $G=c=1$ through most of the paper.} Thus we have $n_{\mu} = -N\,\partial_{\mu}\tau$, where $N$ will later correspond to the ordinary lapse in adapted 3+1 coordinates and reads
\begin{equation}\label{N}
	N = \frac{1}{\sqrt{-g^{\mu\nu}\partial_{\mu}\tau\partial_{\nu}\tau}}\; .
\end{equation} 
The vector $n_\mu$ is invariant by reparametrization $\tau\rightarrow\tilde{\tau}(\tau)$ where $\tilde{\tau}(\tau)$ denotes any strictly increasing function. We also define the projector on the constant-$\tau$ hypersurfaces by $\gamma_{\mu\nu} = g_{\mu\nu}+n_{\mu}n_{\nu}$, and as usual the associated projected covariant derivative $D_{\mu}$ (for instance $D_\mu=\gamma_\mu^{\ph{\mu}\nu}\partial_\nu$ when acting on scalars; see Appendix \ref{appA} for more notation). The acceleration of the congruence orthogonal to the constant-$\tau$ hypersurfaces is given by $a^{\mu} = n^{\nu}\nabla_{\nu}n^{\mu}$ or, equivalently, in terms of the projected spatial derivative operator,
\begin{equation}\label{amu}
        a_{\mu} = D_{\mu} \ln N\; .
\end{equation} 
We will see in Section \ref{secIV} that, in the non-relativistic limit and in adapted coordinates, the spatial components of the acceleration reduce to minus the ordinary gravitational acceleration. Hence observers which would follow the congruence in their motion with $4$-velocity $n^{\mu}$ are essentially un-accelerated observers in coordinates adapted to the constant-$\tau$ hypersurfaces. Thereafter we will call such observers the \textit{fiducial} observers.

In the present paper, motivated by the construction of a relativistic MOND theory modifying gravity in a regime of small accelerations, we investigate a specific extension of the Einstein-Hilbert action for gravity by a function of the norm of the (space-like) acceleration \eqref{amu}, defined by 
\begin{equation}\label{a}
        a = \sqrt{\gamma_{\mu\nu}\,a^{\mu}a^{\nu}}\; .
\end{equation} 
In keeping with MOND we shall assume that such an extension becomes effective only when gravity is weaker than the MOND scale $a_0$ (see Section \ref{secIV}). Well above $a_0$, but still remaining purely classical, we will assume that the standard general relativity with a cosmological constant is recovered.

We consider a Lagrangian density $\mathcal{L}[g_{\mu\nu},\tau,\Psi]$ depending on three dynamical fields, the metric $g_{\mu\nu}$, the Khronon field $\tau$ associated with the time foliation, and the matter fields collectively denoted by $\Psi$. Specifically we choose
\begin{equation}\label{L4D}
	\mathcal{L} = \frac{\sqrt{-g}}{16\pi}\Bigl[ R - 2f(a)\Bigr] + \mathcal{L}_\text{m}[g_{\mu\nu},\Psi] \;,
\end{equation}
where $R$ is the curvature scalar, $\mathcal{L}_\text{m}$ denotes the Lagrangian of the matter fields, and $f(a)$ is our postulated modification, in the form of an unspecified function of the norm of the acceleration \eqref{a}. Only the asymptotic behaviour of $f(a)$ when $a\to\infty$ and when $a\to 0$ will be specified in Section \ref{secIV}. The factor in front of $f(a)$ is chosen so that a constant term in $f(a)$ will correspond to the usual definition of the cosmological constant.

The specific choice for the Lagrangian \eqref{L4D} corresponds to one particular term investigated in MOND-motivated non-canonical versions of Einstein-{\ae}ther theories: see for instance the $c_4$-term in the equation (30) of \cite{HZL08}. This term, among the simplest kinetic terms considered in Einstein-{\ae}ther theories, is sufficient for our purpose of reproducing the MOND dynamics. It is also the only one to cancel out in an homogeneous and isotropic Universe (see \textit{e.g.} \cite{ZZ10}). Note that the expression \eqref{amu} we employ here for the acceleration is valid only in the hypersurface-orthogonal case, and implies for instance that $D_\mu a_\nu=D_\nu a_\mu$.

The function $f(a)$ is the only term depending on the Khronon field $\tau$ in the action \eqref{L4D}. In adapted coordinates, this term will be interpreted as a Lorentz invariance breaking term associated with the preferred frame effect. Here we assume that the matter Lagrangian $\mathcal{L}_\text{m}$ is standard, \textit{i.e.} given by the standard universal coupling to the metric $g_{\mu\nu}$, without dependence on the $\tau$-field. Apart from that, we do not make any restriction on the Lagrangian of the matter fields.

Varying the action with respect to the metric yields a modified Einstein field equation,
\begin{equation}\label{EE}
	G^{\mu\nu}+f(a)g^{\mu\nu} + 2n^{\mu}n^{\nu}\nabla_{\rho}\bigl[\chi(a) a^{\rho}\bigr] - 2\chi(a) a^{\mu}a^{\nu} = 8\pi \,T^{\mu\nu} \; ,
\end{equation}
whose trace gives
\begin{equation}\label{EEtrace}
	R - 4 f(a) + 2 \nabla_{\mu}\bigl[\chi(a) a^{\mu}\bigr] + 2\chi(a) a^2 =  - 8\pi \,T \; .
\end{equation}
Here $G^{\mu\nu}=R^{\mu\nu}-\frac{1}{2}g^{\mu\nu}R$ is the Einstein tensor, $T^{\mu\nu}=(2/\sqrt{-g})\delta\mathcal{L}_\text{m}/\delta g_{\mu\nu}$ is the stress-energy tensor of the matter fields, and we have introduced the convenient notation
\begin{equation}\label{chi}
	 \chi(a)=\frac{f'(a)}{2a}\; ,
\end{equation}
with $f'(a)=\ud f/\ud a$. Such notation is motivated by the non-relativistic MOND limit, in which $\chi(a)$ will appear as a gravitational analogue of the electric susceptibility of the (phantom) dark matter medium in the sense of \cite{B07mond}.

Varying next with respect to the scalar field $\tau$ leads to an independent equation, called the $\tau$-equation in the following, and given by (see \cite{Jacobson10})\footnote{In the case where the {\ae}ther vector is not hypersurface orthogonal, the quantity defined by $K^{\mu\nu} = \nabla^{\mu}n^{\nu} + n^{\mu}a^{\nu}$ is not symmetric anymore (and does not warrant to be called an extrinsic curvature). Varying in this case the action with respect to $n_{\mu}$, one obtains instead of the scalar equation \eqref{Teq0} the \textit{vectorial} equation
$$n^\nu\nabla_{\nu}(\chi a^{\mu})-\chi a^2 n^\mu - \chi\left(K^{\mu\nu}-\gamma^{\mu\nu}K\right)a_\nu = 0\;,
$$
which has three independent components. Varying with respect to the metric, one also gets a different modified Einstein equation \eqref{EE}, with one additional term in the left-hand side given by $- 4\chi a_{\rho}n^{(\mu}A^{\nu)\rho}$ where we pose $A^{\mu\nu}=K^{\mu\nu}-K^{\nu\mu}$.}
\begin{equation}\label{Teq0}
	\nabla_\mu\Bigl[N\Bigl(n^\nu\nabla_{\nu}(\chi a^{\mu})-\chi a^2 n^\mu - \chi\left(K^{\mu\nu}-\gamma^{\mu\nu}K\right)a_\nu\Bigr)\Bigr] = 0\; ,
\end{equation}
where the extrinsic curvature tensor, which is symmetric, is defined by $K^{\mu\nu} = \nabla^{\mu}n^{\nu} + n^{\mu}a^{\nu}$ (with trace given by $K=\nabla_{\mu}n^{\mu}$). An alternative, simpler form of this equation, reads as
\begin{equation}\label{Teq}
	\nabla_{\mu}\Bigl[ n^{\mu}\nabla_{\nu}(\chi a^{\nu})\Bigr] = \frac{\dot{f}}{2} + \chi \,a^{\mu}a^{\nu}K_{\mu\nu}\; ,
\end{equation}
where we denote $\dot{f}=n^{\nu}\nabla_{\nu}f$. Note that this $\tau$-equation is of fourth order in derivatives of the Khronon field $\tau$. However, as will become clear in Section \ref{secIII}, if we select a coordinate system for which $t=\tau$, the $\tau$-equation will be equivalent to an equation of first order only in time derivatives of geometrical quantities.

When we rephrase the model by interpreting $\tau$ as a matter field, \textit{i.e.} rewriting the modified Einstein equation in the form $G^{\mu\nu} = 8\pi (T^{\mu\nu}+T^{\mu\nu}_{\tau})$, where the stress-energy tensor of this matter field (say a ``dark energy'' fluid) reads
\begin{equation}\label{Tlambda}
	T^{\mu\nu}_{\tau} = - \frac{1}{8\pi}\Bigl[ f g^{\mu\nu} + 2n^{\mu}n^{\nu}\nabla_{\rho}(\chi a^{\rho}) - 2\chi a^{\mu}a^{\nu}\Bigr] \; ,
\end{equation}
the $\tau$-equation is equivalent to the conservation of this associated energy-momentum tensor, $\nabla_{\nu}T^{\mu\nu}_{\tau} = 0$. Thus, given the modified Einstein equation \eqref{EE} and the contracted Bianchi identity $\nabla_{\nu}G^{\mu\nu} \equiv 0$, we see that the $\tau$-equation is in fact equivalent to the conservation of the matter energy-momentum tensor $\nabla_{\nu}T^{\mu\nu}=0$, which is itself the consequence of the matter field equations and the scalarity of the action (see \cite{Jacobson10} for discussion). In Appendix \ref{appB}, we will give the canonical decomposition of the stress-energy tensor \eqref{Tlambda}. 

\section{Coordinates adapted to the foliation}
\label{secIII}

In this Section, we express the model in an equivalent 3+1 formalism, choosing a coordinate system adapted to the foliation. Note that this re-writing is not possible in a generic Einstein-{\ae}ther theory like in \cite{ZFS07,HZL08}. The time coordinate will coincide with the Khronon field: $t=\tau$ and thus $n_{\mu}=(-N,\mathbf{0})$ with $N=1/\sqrt{-g^{00}}$. Introducing the shift vector $N_{i}=g_{0i}$ and the spatial metric $\gamma_{ij}=g_{ij}$, we have the usual 3+1 form for the metric, $N$ playing the role of the lapse:
\begin{equation}
	ds^2 = -(N^2-N_{i}N^{i})\ud t^2 + 2N_{i}\ud t \ud x^{i} + \gamma_{ij}\ud x^{i}\ud x^{j} \; .
\end{equation}
In particular $n^\mu=\frac{1}{N}(1,-N^i)$. All spatial (Latin) indices are raised and lowered using the spatial metric $\gamma_{ij}$ and its inverse $\gamma^{ij}=g^{ij}+N^{i}N^{j}/N^2$ (such that $\gamma^{ij}\gamma_{jk}=\delta^{i}_{k}$) which coincides with the spatial components of the 4-dimensional projector $\gamma^{\mu\nu}$. 

The 4-dimensional Lagrangian \eqref{L4D} now becomes, after discarding an irrelevant total divergence term (with $\sqrt{-g} = N\sqrt{\gamma}$),
\begin{equation}\label{L3D} 
	\mathcal{L} = \frac{\sqrt{\gamma}}{16\pi}N\Bigl[ \mathcal{R}+K_{ij}K^{ij}-K^{2}-2f(a)\Bigr] + \mathcal{L}_\text{m}[N,N_{i},\gamma_{ij},\Psi] \;.
\end{equation}
Here $\mathcal{R}$ denotes the 3-dimensional scalar curvature associated with the spatial metric $\gamma_{ij}$. The projected derivative operator reduces to the spatial covariant derivative $D_{i}$ associated with $\gamma_{ij}$, and the extrinsic curvature takes the usual expression in adapted coordinates,
\begin{equation}\label{Kij} 
	K_{ij} = \frac{1}{2N}\bigl( \partial_{t}\gamma_{ij}-D_{i}N_{j}-D_{j}N_{i} \bigr) \; .
\end{equation}
The acceleration gets now the 3-dimensional expression
\begin{equation}\label{ai} 
	 a^{i}=D^{i}\ln N\; ,\qquad a=\sqrt{\gamma_{ij}a^{i}a^{j}}\; ,
\end{equation}
where $D^i=\gamma^{ij}D_j$ (notice that $a^0=0$).

The $3d$ theory \eqref{L3D} is equivalent to the previous $4d$ theory \eqref{L4D}, as is explicitly checked in Appendix \ref{appA} at the level of the field equations. However the spirit is different, since the $\tau$-field disappears from \eqref{L3D} and is now hidden in the time coordinate. For instance, the acceleration's norm $a$ was a $\tau$-dependent quantity in \eqref{L4D}, but it depends now only on the lapse $N$ which is a geometrical degree of freedom. 

Hence one may interpret the new $a$-dependent term in the 3+1 Lagrangian \eqref{L3D} as a Lorentz invariance violating term. While the $4d$ theory \eqref{L4D} is generally covariant, \textit{i.e.} invariant under the full group of $4d$ diffeomorphisms
\begin{equation}\label{diffeo4d} 
	 \tilde{x}^\mu=\tilde{x}^\mu(x^\nu)\; ,
\end{equation}
[with $\tau$ being a scalar field, \textit{i.e.} $\tilde{\tau}(\tilde{x})=\tau(x)$], the $3d$ formulation \eqref{L3D} is only invariant under the subgroup of diffeomorphisms leaving invariant the preferred time foliation, namely
\begin{subequations}\label{diffeo3d} 
\begin{align}
	 \tilde{t} &= \tilde{t}(t)\; ,\\ \tilde{x}^i &= \tilde{x}^i(x^j,t)\; .
\end{align}\end{subequations}
Under this subgroup of diffeomorphisms the lapse and shift transform as
\begin{subequations}\label{transf3d} 
\begin{align}
	 \tilde{N} &= \frac{\ud t}{\ud\tilde{t}}\,N\; ,\label{transfN}\\ \tilde{N}_i &= \frac{\ud t}{\ud\tilde{t}}\,\frac{\partial x^j}{\partial\tilde{x}^i}\,N_j+\frac{\partial x^j}{\partial\tilde{t}}\,\frac{\partial x^k}{\partial\tilde{x}^i}\,\gamma_{jk}\; ,
\end{align}\end{subequations}
while the spatial metric transforms like a spatial tensor. Note that \eqref{transfN} corresponds in the $4d$ formalism (such that $t=\tau$) to a reparametrization of the time foliation by $\tilde{\tau}(\tau)=\tilde{t}(t)$.

The broken diffeomorphism invariance \eqref{diffeo3d} has recently been used in the context of quantum gravity to build a power-counting renormalizable quantum gravity at high energy \cite{Horava09,BPS10,BPS11}. As is known, the full diffeomorphism invariance can be restored in the $4d$ formulation at the price of an additional structure. This is the role of the Khronon field $\tau$, which appears to be the St\"uckelberg field \cite{Stueck38} associated with the broken diffeomorphism symmetry.\footnote{A well-known example of Khronon field is the one introduced in the Cartan theory of gravity which is the covariantization of Newton's theory \cite{Cartan23,Cartan24,Traut63}.} 

To write the field equations in the most convenient way we use the following notations as a shortcut for describing the matter part of the field equations:
\begin{subequations}\label{fluid}
\begin{align}
	\varepsilon &= -\frac{1}{\sqrt{\gamma}}\frac{\delta\mathcal{L}_{m}}{\delta N} \;,\label{fluida}\\
	J^{i} &= \frac{1}{\sqrt{\gamma}}\frac{\delta\mathcal{L}_{m}}{\delta N_{i}}\;, \\
	\mathcal{T}^{ij} &= \frac{2}{N\!\sqrt{\gamma}}\frac{\delta\mathcal{L}_{m}}{\delta \gamma_{ij}}\;.
\end{align}
\end{subequations}
These quantities reduce to the usual notions of the energy density, current density and spatial stresses in the case of a perfect fluid. However they apply to any kind of matter fluid or field. The variation of the Lagrangian \eqref{L3D} with respect to the lapse $N$ yields a modified Hamiltonian constraint equation, namely
\begin{equation}\label{ham}
	\mathcal{R} + K^2 - K_{ij}K^{ij} - 2f + 4\chi a^2 + 4D_{i}(\chi a^{i}) = 16\pi \,\varepsilon\;,
\end{equation}
where we recall that both $f(a)$ and $\chi(a)=f'(a)/(2a)$ are functions of the norm of the acceleration given in \eqref{ai}. The variation with respect to the shift $N_{i}$ yields the momentum constraint equation, \textit{i.e.}
\begin{equation}\label{constr}
	D_{j}\left( K^{ij}-\gamma^{ij}K \right) = - 8\pi J^{i}\;.
\end{equation}
This equation is unchanged with respect to general relativity, since the term we have added in the Lagrangian does not depend on the shift. Finally, varying with respect to the spatial metric $\gamma_{ij}$ gives the evolution equation
\begin{align}\label{evol}
\mathcal{G}^{ij} &+ \frac{1}{N}D_t\left(K^{ij}-\gamma^{ij}K\right)+ \frac{2}{N}D_{k}\left[N^{(i}(K^{j)k}-\gamma^{j)k}K)\right]\nonumber \\
& + 2 K^{ik}K^{j}_{\ph{j}k}-K K^{ij}- \frac{1}{2}\gamma^{ij}\left(K^{kl}K_{kl}+K^2\right) \nonumber \\
	& - \frac{1}{N} \left(D^iD^jN - \gamma^{ij} D_kD^k N\right) - 2\chi a^ia^j + f \gamma^{ij} = 8\pi \mathcal{T}^{ij}\,,
\end{align}
where $\mathcal{G}^{ij}=\mathcal{R}^{ij}-\frac{1}{2}\gamma^{ij}\mathcal{R}$ is the 3-dimensional Einstein tensor, and the convenient notation $D_t=\partial_t-N^kD_k$ is used. We gave here the full equation for completeness, but the only difference with general relativity lies in the two last acceleration-dependent terms with $f(a)$ and $\chi(a)$. The trace of \eqref{evol} gives
\begin{equation}\label{evoltrace}
	 \mathcal{R} + \frac{4}{N}D_t K + 3 K_{ij}K^{ij} + K^2 - \frac{4}{N} D_iD^i N - 6 f + 4\chi a^2 = -16\pi\left( \mathcal{T} + \frac{2}{N}N_i J^i \right) \;,
\end{equation}
where $\mathcal{T}=\gamma_{ij}\mathcal{T}^{ij}$, and we also used the momentum constraint equation \eqref{constr}. By eliminating $\mathcal{R}$ using the Hamiltonian constraint \eqref{ham}, one obtains an important equation, which plays the role of a modified Poisson-type equation for the gravitational field,
\begin{equation}\label{modPoisson}
	D_{i}\left[(1+\chi)a^i\right] + f + a^2 - \frac{1}{N}D_t K - K^{ij}K_{ij} = 4\pi\left(\varepsilon + \frac{2}{N}N_iJ^i + \mathcal{T}\right)\; .
\end{equation}

We check in Appendix \ref{appA} the compatibility of the field equations in the $3d$ and $4d$ representations, that are obtained by varying with respect to different degrees of freedom, \textit{i.e.} $g_{\mu\nu}$, $\tau$ and $\Psi$ in the $4d$ formulation, and $N$, $N_i$, $\gamma_{ij}$ and $\Psi$ in the $3d$ formulation. Starting with the $4d$ Einstein equation \eqref{EE} and applying the projection procedure which allows to get the 3+1 equations with the help of the Gauss-Codazzi relations, we get the same 3+1 equations as obtained by varying directly the action expressed in adapted cordinates, namely \eqref{ham}--\eqref{evol}. In other words, starting with the action in 3+1 coordinates and varying the non-covariant degrees of freedom, or varying the $4d$ covariant action and then projecting the equations obtained in adapted 3+1 coordinates, yield the same field equations. 

A related question is the fate of the Khronon field equation \eqref{Teq} when switching to adapted 3+1 coordinates. Since $\tau$ becomes the time coordinate, we no longer vary the action with respect to it, and the three equations \eqref{ham}--\eqref{evol} do not manifestly contain the $\tau$-field.
In adapted 3+1 coordinates the equation \eqref{Teq} becomes\footnote{Note that in 3+1 adapted coordinates, we have $\nabla_{\mu}(\chi a^{\mu}) = D_{i}(\chi a^{i}) + \chi a^2$.}
\begin{equation}\label{Teq3D}
D_t\biggl[D_{i}(\chi a^{i}) + \chi a^2 - \frac{f}{2}\biggr] + N K \Bigl(D_{i}(\chi a^{i}) + \chi a^2\Bigr) - N \chi \,a^{i}a^{j}K_{ij} = 0 \; ,
\end{equation}
where $D_t=\partial_t-N^kD_k$. We substitute the term $D_{i}(\chi a^{i})$ by using the Hamiltonian constraint \eqref{ham} and perform a series of transformations with the help of the other equations \eqref{constr}--\eqref{evol}. A long calculation, in which one uses the useful relation
\begin{equation}\label{relation} 
D_iD_j\Bigl[N\left(K^{ij}-\gamma^{ij}K\right)\Bigr] = \frac{1}{2}D_t\mathcal{R} + N K^{ij}\mathcal{R}_{ij}\; ,
\end{equation}
then shows that \eqref{Teq3D} is equivalent to the continuity equation for the matter field, \textit{i.e.} the evolution equation of the matter energy density $\varepsilon$ defined in \eqref{fluida}, which is given by
\begin{equation}\label{cont}
D_t\varepsilon + N K \varepsilon + D_i(N J^i) + J^i D_iN + 2K_{ij}N^iJ^j+N\mathcal{T}^{ij}K_{ij} = 0 \; .
\end{equation}
This equation is the $0$-th component of the conservation law for the matter field, \textit{i.e.} $\nabla_{\nu}T^{0\nu} = 0$. Furthermore, another long computation shows that, given the $3d$ contracted Bianchi identity $D_j\mathcal{G}^{ij} \equiv 0$, and the equations \eqref{ham}-\eqref{evol}, the Euler equation of the matter field is satisfied, \textit{i.e.} the evolution equation of the current $J^i$ given by
\begin{equation}\label{euler}
D_t J^i +N K J^i + D_j\left[N \mathcal{T}^{ij}+N^iJ^j+N^jJ^i\right] + \varepsilon D^iN + J^jD_jN^i+2N J^j K^i_j = 0 \; .
\end{equation}
This corresponds to the spatial component of the matter conservation $\nabla_{\nu}T^{i\nu} = 0$, once a term is substituted using the continuity equation \eqref{cont}. These checks are the transposition in the 3+1 formalism of the fact that the $\tau$-equation \eqref{Teq} is implied by the Einstein field equation \eqref{EE} (see \cite{Jacobson10} and Section \ref{secII}).

\section{Non-relativistic limit}
\label{secIV}

We now investigate the non-relativistic limit of the model, which will determine the required behaviour of the function $f(a)$ in order to recover a MOND-like modification of the Poisson equation in the relevant weak field regime. In this Section we restore the $G$ and $c$ factors; in the field equations \eqref{evol} for instance, the right-hand side must be multiplied by $G /c^4$ so that the stress-energy tensor has the dimension of an energy density. 

We consider a matter system isolated from the rest of the Universe and at rest with respect to the preferred frame, in the sense that the fiducial observers have a coordinate velocity $w^i=n^i/n^0=-N^i$ which is of small post-Newtonian order. We make the usual post-Newtonian ansatz on the metric components generated by the system:
\begin{subequations}\label{PNmetric}
\begin{align}
	N &= 1 + \frac{\phi}{c^2} + \mathcal{O}\left( 4 \right)\;, \\
	N_{i} &= \mathcal{O}\left( 3 \right)\;, \\
	\gamma_{ij} &= \delta_{ij}\left(1 - \frac{2\psi}{c^2}\right) + \mathcal{O}\left( 4 \right)\;,\label{PNmetricc}
\end{align}
\end{subequations}
where $\mathcal{O}(n)$ denotes a small remainder term of order $\mathcal{O}\left( c^{-n} \right)$. We introduce in \eqref{PNmetric} two \textit{a priori} different Newtonian-like potentials $\phi$ and $\psi$ for the lapse and the spatial metric, but we shall check that they are actually identical like in general relativity.

The acceleration \eqref{ai} of the fiducial observers reduces to the Newtonian acceleration of the $\phi$-potential,
\begin{equation}\label{aN}
	 \bm{a} = \frac{1}{c^2}\vecnab \phi + \mathcal{O}\left( 4 \right) \;.
\end{equation}
We use bold-face notation to represent ordinary three-dimensional Euclidean vectors, \textit{e.g.} $\bm{a}=(a^i)$. With our convention, $\bm{a}$ has the dimension of the inverse of a length. From \eqref{Kij} we see that the extrinsic curvative is a small post-Newtonian quantity, of order $K_{ij} = \mathcal{O}( 3 )$. Finally we assume that the matter source is non-relativistic, so that its energy-momentum tensor is dominated by the Newtonian rest mass density denoted $\rho$:
\begin{equation}\label{nonrel1}
	\varepsilon = \rho\,c^2 + \mathcal{O}\left( 0 \right) \;,
\end{equation}
with comparatively negligible current and stress densities,
\begin{subequations}\label{nonrel2}
\begin{align}
	J^{i} &= \rho v^i + \mathcal{O}\left( 2 \right)\;, \\
	\mathcal{T}^{ij} &= \rho v^i v^j + \mathcal{O}\left( 2 \right)\;.
\end{align}
\end{subequations}
Then the equations \eqref{cont} and \eqref{euler} reduce to the standard Newtonian continuity and Euler equations (with $\phi$-potential).

We first show that the two Newtonian potentials $\phi$ and $\psi$ are equal. This follows from the trace of the evolution equation given by \eqref{evoltrace}. The function $f(a)$ scales with $a^2$ where the dimensionful acceleration is of order $a=\mathcal{O}(2)$ from \eqref{aN}. This means that the post-Newtonian order of this function is $f(a) = \mathcal{O}(4)$. Note that since $f(a)$ will contain a cosmological constant $\Lambda$ (see \textit{e.g.} \eqref{lambdaMond} below), to be consistent the post-Newtonian order of the cosmological constant must also be $\Lambda = \mathcal{O}(4)$. Inserting into \eqref{evoltrace} the previous estimate $f(a) = \mathcal{O}(4)$ together with $K_{ij} = \mathcal{O}(3)$, and using the fact that at the dominant order we have $\mathcal{R} = \frac{4}{c^2} \Delta \psi + \mathcal{O}( 4 )$, we readily obtain $\Delta (\psi - \phi) = \mathcal{O}(2)$. For a regular isolated matter source the only solution is
\begin{equation}\label{equal}
	 \psi = \phi + \mathcal{O}\left( 2 \right) \; .
\end{equation}
The equality of the two potentials is very important for the viability of the theory as an alternative to dark matter (see \cite{Bekrev} for a review). It implies that the light deflection and the gravitational lensing will be given by the same formula as in general relativity, but with a single potential $\phi = \psi$ obeying the modified Poisson equation (given by \eqref{modPoissonNR} below). Unlike in TeVeS \cite{Bek04,Sand05}, there is no need in the present theory to invoke a disformal coupling of the photon field to gravity in order to get the light deflection viable. The property \eqref{equal} has been proved in \cite{ZFS07} and is a generic feature of non-canonical Einstein-{\ae}ther theories. 

We now turn to the equation satisfied by the Newtonian potential $\phi$. Clearly, from the form of the metric \eqref{PNmetric}, the ordinary baryonic matter feels the Newtonian gravitational field $\bm{g}=-\vecnab \phi$. From \eqref{aN} we have $\bm{a}=-\bm{g}/c^2$ at the leading order. In the non-relativistic limit, and for the non-relativistic matter source satisfying \eqref{nonrel1}--\eqref{nonrel2}, the equation \eqref{modPoisson} becomes
\begin{equation}\label{modPoissonNR}
\vecnab \cdot \Bigl[ \left( 1+ \chi \right) \vecnab \phi
  \Bigr] = 4\pi G \rho + \mathcal{O}\left( 2 \right)\;,
\end{equation}
which exactly reproduces the MOND equation \eqref{MOND} for the gravitational field $\bm{g}=-\vecnab \phi$. The MOND function is simply given by $\mu = 1+\chi$, and $\chi$ can be interpreted as a ``susceptibility'' coefficient for the modified Poisson equation. Using the fact that $a=g/c^2$ at the leading order, the latter correspondence between $\mu(g/a_0)$ and the function $\chi(a) =f'(a)/(2a)$ will tell us what are the constraints to be imposed on the function $f(a)$ in the initial Lagrangian. 

To recover the Newtonian regime when $a\gg a_0$ (\textit{i.e.} formally when $a\to\infty$), this function should tend toward a constant. More precisely, we shall recover general relativity with a cosmological constant which is the limiting value of $f(a)$, say 
\begin{equation}\label{lambdaGR}
f(a) = \Lambda_\infty ~~\text{when}~~ a\to\infty\;.
\end{equation}
On the other hand, to recover MOND asymptotically, we require that when $a\to 0$ ($a$ has the dimension of $[\text{length}]^{-1}$),
\begin{equation}\label{lambdaMond}
f(a) = \Lambda_0 - a^2 + \frac{2c^2}{3 a_0}a^3+ \mathcal{O}\left(\frac{c^4 a^4}{a_0^2}\right)\;,
\end{equation}
where $\Lambda_0$ is constant, and we recall that the MOND scale is measured to the value $a_0\simeq 1.2\times 10^{-10}\,\text{m}/\text{s}^2$ \cite{SandMcG02}. We see that the term $\propto a^2$ in the expansion \eqref{lambdaMond} is similar to one of the terms proposed in the ``healthy'' extension of Ho\v{r}ava gravity \cite{BPS10,BPS11}. Using the notation of \cite{BPS10,BPS11} the coefficient in front of this term in the action is $\alpha=2$. Beware, however, that this term here is only the first term in a weak-field expansion. 

\textit{A priori}, the cosmological constant $\Lambda_\infty$ appearing in the general relativistic limit \eqref{lambdaGR} could be different from the cosmological constant $\Lambda_0$ in the MOND regime \eqref{lambdaMond}. Notice that in an homogeneous and isotropic Universe, making the $\tau$-field to coincide with cosmic time, $a$ vanishes identically and the equivalent energy-momentum tensor \eqref{Tlambda} reduces to a cosmological constant $f(0)=\Lambda_0$. However, the precise cosmological implications of the model would need a study at the level of perturbations, which is left for future work (see \cite{SMFB06,LMB07,Sk08,ZZ10} for cosmological studies in modified gravity theories). For simplicity, we provisionally assume here that $\Lambda_\infty$ and $\Lambda_0$ are equal and we denote their common value by $\Lambda$. 

Note that in the MOND regime at least, the post-Newtonian order of magnitude $\Lambda=\mathcal{O}(4)$ we have assumed for the cosmological constant is natural. Indeed, because of the numerical coincidence between $a_0$ and the acceleration scale $a_{\Lambda}\sim \sqrt{\Lambda}$ associated with the cosmological constant \cite{Milgrev}, $\Lambda$ scales like $a_0^2/c^4$ so that we indeed have $\Lambda=\mathcal{O}(4)$. In this picture we are considering that $a_0$ represents a new fundamental constant independent from $c$ and $G$, presumably coming from some new (unknown) fundamental physics. Therefore, factorizing out $\Lambda$ in \eqref{lambdaMond} would yield a Taylor expansion in terms of the small dimensionless parameter $c^2 a/a_0$, with coefficients expected to be of the order of one.

The coincidence between $a_0$ and $a_\Lambda$ can be made completely explicit in the following example. A function fulfilling all the above requirements is the ``Planckian'' distribution
\begin{equation}\label{planck}
f(a) = \Lambda - \frac{a^3}{a_\Lambda\bigl[e^{a/a_\Lambda}-1\bigr]}\;,
\end{equation}
where the ``temperature'' $a_\Lambda$ (in fact $k T_\Lambda = \hbar c a_\Lambda$) is that of the horizon of the de Sitter Universe associated with the positive cosmological constant, namely \cite{GiHaw77}
\begin{equation}\label{alambda}
a_{\Lambda} = \frac{1}{2\pi}\sqrt{\frac{\Lambda}{3}}\;.
\end{equation}
By comparing the expansion when $a\to 0$ of this function with \eqref{lambdaMond} we deduce that $a_0$ would be related to the cosmological constant $\Lambda$ by
\begin{equation}\label{coincidence}
a_0=\frac{4 a_\Lambda c^2}{3}\;.
\end{equation}
The numerically value of $a_0$ would be extremely close to the actual one: indeed we have $c^2 a_\Lambda\simeq 0.9\times 10^{-10}\,\text{m}/\text{s}^2$ so that $a_0\simeq 1.2\times 10^{-10}\,\text{m}/\text{s}^2$, in very good agreement with observations. However, the function \eqref{planck} is only an example without physical justification.

Note that we can also retrieve the deep MOND regime directly at the level of the Lagrangian \eqref{L3D}. Indeed we have, up to a total divergence, $\sqrt{\gamma} N \mathcal{R} = -\frac{2}{c^4}\vert\vecnab \psi\vert^2 + \mathcal{O}(6)$, in terms of the potential $\psi$ parametrizing the spatial metric \eqref{PNmetricc}. But because the two potentials $\psi$ and $\phi$ are equal, we see that this term will be cancelled by the term $2a^2 = \frac{2}{c^4}\vert\vecnab \phi\vert^2 + \mathcal{O}(6)$ coming from the expansion of $f(a)$ as given by \eqref{lambdaMond}. Therefore we find that the Newtonian behaviour is suppressed, and it remains the next term in the expansion which is proportional to $a^3$ and immediately gives the MOND Lagrangian
\begin{equation}
	\mathcal{L}_\text{MOND} = -\frac{1}{12\pi G a_{0}} \vert\vecnab \phi\vert^3 - \rho \,\phi\;,
\end{equation}
whose variation with respect to $\phi$ reproduces the deep-MOND equation
\begin{equation}
	\vecnab \cdot \biggl( \frac{\vert\vecnab \phi\vert}{a_0}\vecnab \phi \biggr) = 4 \pi G \rho \;.
\end{equation}
The cancellation of the Newtonian term in the Lagrangian by a ``counter-term'' $\propto a^2$ and its replacement by the next-order cubic term $\propto a^3$ is similar to what happens in other approaches, see \textit{e.g.} \cite{ZFS07} and equations (67)--(69) in \cite{DEFW11}.

Thus we have recovered the MOND equation in the appropriate regime. Beware however that the analysis presented in this Section relies on the fact that we choose coordinates adapted to the foliation. If we come back to a $4d$ formulation with the $\tau$-field being a dynamical field, we still may make the hypothesis \eqref{PNmetric} for the metric generated by a system at rest in a given frame, but the expression \eqref{aN} becomes
\begin{equation}\label{aNmod}
	\bm{a} = \frac{1}{c^2}\left[ \frac{\ud\bm{w}}{\ud t} + \vecnab \phi \right]+ \mathcal{O}\left( 4 \right) \; ,
\end{equation}
with $\bm{w}=\bm{n}/n^0$ being the coordinate velocity of fiducial observers in that frame and $\ud\bm{w}/\ud t$ being their coordinate acceleration. The fiducial observers will follow a non-geodesic motion governed by the $\tau$-equation, thus $\bm{a}\neq \bm{0}$ in general, but the first term in \eqref{aNmod} will appear as a preferred-frame effect. Note that this effect depends on the acceleration and occurs already at Newtonian order. In the strong field regime where $|\vecnab \phi| \gg a_0$ (Solar system, binary pulsars, \textit{etc.}), the presence of this term should not change the fact that $a \gg a_0/c^2$, so the modification of the dynamics with respect to general relativity should be suppressed. In the weak-field regime however, for instance in the outskirts of galaxies, this term could lead to a sizeable effect. We would need to investigate the complicated dynamics driven by the $\tau$-equation to get a precise answer. However if we assume that the preferred frame essentially coincides with the frame of the large scale structure or the cosmic microwave background, this coordinate acceleration of fiducial observers in a galaxy-centered frame becomes (minus) the acceleration of the galaxy with respect to the CMB frame, which is typically of the order of magnitude of $a_0/100$ (see \textit{e.g.} \cite{FBZ08}). In this case the term $\ud\bm{w}/\ud t$ in \eqref{aNmod} is not expected to significantly impact the fit to MOND phenomenology.   
 
\section{Conclusion}
\label{secV}

We have proposed a relativistic modified gravity model, based on a preferred 3+1 space-time foliation, reproducing the phenomenology of MOND \cite{SandMcG02} in the weak-field limit. The modification with respect to general relativity consists in adding to its ordinary Lagrangian a function of the norm of the acceleration of the congruence associated with this foliation. We investigated two different points of view on this theory: In the first, we introduced a scalar field called the Khronon defining the foliation by its constant-value hypersurfaces, keeping a full $4$-dimensional covariant formalism, while in the second, we wrote the theory in a 3+1 fashion, in a frame where the time coordinate coincides with the Khronon field and where the Lagrangian is no longer manifestly Lorentz invariant. The spirit of our approach is similar as that of recent attempts at building a consistent quantum theory of gravity \cite{Horava09,BPS10,BPS11}, but here the Lorentz invariance violation occurs at low energy, and our motivation is purely classical. We showed the equivalence between the $3d$ and $4d$ formulations of the model, and gave the requirements on the function initially introduced in the Lagrangian to recover MOND in the non-relativistic approximation.

We leave to future work some important questions. First, it is easy to see that in an homogeneous and isotropic background, our model simply reduces to the addition of a cosmological constant. However the cosmological implications of this model in a perturbative regime have not been studied yet. We have in mind the viability of the model when faced to cosmological observations, notably the anisotropies of the cosmic microwave background and the structure formation (see \cite{SMFB06,LMB07,Sk08,ZZ10} for general investigations of the cosmology in TeVeS and Einstein-{\ae}ther theories). 

A second question is that of the viability regarding Solar system tests and binary pulsars data. Although it should recover general relativity plus a cosmological constant in the strong field regime appropriate to these systems, see \eqref{lambdaGR},\footnote{For instance, the MOND transition radius for the Sun is $\sqrt{G M_\odot/a_{0}} \simeq 7100 \text{AU}$.} our model includes a preferred frame effect which we would like to quantify precisely, using various hypothesis for the motion of the Solar system with respect to the preferred frame. (See \cite{FJ06,Sagi09} for the computation of the preferred frame post-Newtonian parameters in TeVeS and canonical Einstein-{\ae}ther.) More importantly, the preferred frame effect should also be computed in the MOND regime, using various forms for the MOND function, to see how it affects the usual MOND fit of the flat rotation curves of galaxies; see \eqref{aNmod} and the discussion right after. 

Another question is related to the dynamics of the Khronon field itself [obeying the equation \eqref{Teq}]. Indeed in this model the foliation is dynamical, and the evolution of the Khronon field should be compatible with the preservation of its smoothness and regularity properties allowing a space-time foliation to be built on it.

\section*{Acknowledgements}
It is a pleasure to thank Gilles Esposito-Far\`ese for useful comments. We acknowledge partial support from Agence Nationale de la Recherche (ANR ``Thales'').

\appendix

\section{3+1 decomposition of the modified Einstein equations}\label{appA}

In this Appendix, we check the equivalence of the $4d$ and $3d$ formulations of the theory discussed in Sections \ref{secII} and \ref{secIII}. The material used here is well-known, and the derivation will basically proceed along the same steps as in standard general relativity (see \textit{e.g.} \cite{Wald}), but we choose to present it for completeness.

The projected covariant derivative operator $D_{\mu}$ is defined in the usual way by \textit{e.g.} $D_{\mu}A^{\nu} = \gamma_{\mu}^{\ph{\mu}\rho}\gamma_{\sigma}^{\ph{\sigma}\nu}\nabla_{\rho}A^{\sigma}$ for any vector belonging to the spatial hypersurface, \textit{i.e.} $n_{\nu}A^{\nu}=0$. The spatial Riemann tensor associated with the derivative $D_{\mu}$ is denoted $\mathcal{R}^{\mu}_{\ph{\mu}\nu\rho\sigma}$. The extrinsic curvature tensor $K_{\mu\nu}=\gamma_{\mu}^{\ph{\mu}\rho}\nabla_{\rho}n^{\nu}$ is symmetric on account of the Frobenius theorem $n_{[\mu}\nabla_{\nu}n_{\rho]}=0$ satisfied by the hypersurface-orthogonal vector $n_{\mu}$. The first Gauss-Codazzi relation
\begin{equation}\label{GC1}
	\mathcal{R}_{\mu\nu\rho\sigma} = \gamma_{\mu}^{\ph{\mu}\alpha} \gamma_{\nu}^{\ph{\nu}\beta} \gamma_{\rho}^{\ph{\rho}\gamma} \gamma_{\sigma}^{\ph{\sigma}\delta}R_{\alpha\beta\gamma\delta} + K_{\mu\sigma}K_{\nu\rho} - K_{\mu\rho}K_{\nu\sigma} \; ,
\end{equation}
is contracted with $\gamma^{\mu\rho}\gamma^{\nu\sigma}$ which yields
\begin{equation}\label{GC1contr}
	\mathcal{R} = R + 2n^{\mu}n^{\nu}R_{\mu\nu} + K_{\mu\nu}K^{\mu\nu} - K^2 \; .
\end{equation}
Inserting into \eqref{GC1contr} our modified version of the Einstein equations \eqref{EE}--\eqref{EEtrace}, we obtain the modified Hamiltonian constraint as
\begin{equation}\label{ham1}
	\mathcal{R} + K^2 - K_{\mu\nu}K^{\mu\nu} - 2f + 4\nabla_{\mu}\bigl(\chi a^{\mu}\bigr) = 16\pi \,n^{\mu}n^{\nu}T_{\mu\nu} \; .
\end{equation}
Next, the second Gauss-Codazzi relation, 
\begin{equation}\label{GC2}
	D_{\nu}\left[ K^{\mu\nu}-\gamma^{\mu\nu}K \right] = \gamma^{\mu\nu}n^{\rho}R_{\nu\rho}\; ,
\end{equation}
when projecting the modified Einstein equation along $\gamma^{\mu\nu}n^{\rho}$, leads to
\begin{equation}\label{constr1}
	D_{\nu}\left[ K^{\mu\nu}-\gamma^{\mu\nu}K \right] = 8\pi \gamma^{\mu\nu}n^{\rho}T_{\nu\rho}\; .
\end{equation}
The $f(a)$-dependent terms disappear in this second constraint equation, as expected since it corresponds to the variation of the action with respect to the shift, and our modification does not concern the shift. Using the fact that 
\begin{equation}
	n^{\rho}n^{\sigma}R_{\mu\rho\nu\sigma} = K_{\mu\rho}K_{\nu}^{\ph{\nu}\rho} + \frac{1}{N}D_{\mu}D_{\nu}N - \frac{1}{N}\mathcal{L}_{\ell}K_{\mu\nu} \; ,
\end{equation}
one contraction of the first Gauss-Codazzi relation yields
\begin{equation}
	\mathcal{R}_{\mu\nu} = \gamma_{\mu}^{\ph{\mu}\rho} \gamma_{\nu}^{\ph{\nu}\sigma}R_{\rho\sigma} - \frac{1}{N}\mathcal{L}_{\ell}K_{\mu\nu} + \frac{1}{N}D_{\mu}D_{\nu}N + 2K_{\mu\rho}K_{\nu}^{\ph{\nu}\rho} - K K_{\mu\nu}\; .
\end{equation}
Here $\mathcal{L}_{\ell}$ denotes the Lie derivative with respect to the vector $\ell^\mu = N n^{\mu}$. Projecting the modified Einstein equation on $\gamma_{\mu}^{\ph{\mu}\rho} \gamma_{\nu}^{\ph{\nu}\sigma}$, we find:
\begin{align}\label{evol1}
	\mathcal{G}_{\mu\nu} &+ \frac{1}{N}\bigl( \mathcal{L}_{\ell}K_{\mu\nu} - \gamma_{\mu\nu}\gamma^{\rho\sigma}\mathcal{L}_{\ell}K_{\rho\sigma} \bigr) + \frac{1}{N}\gamma_{\mu\nu} D^{\rho}D_{\rho} N - \frac{1}{N}D_{\mu}D_{\nu}N \nonumber \\
	&+ K K_{\mu\nu} - 2K_{\mu\rho}K_{\nu}^{\ph{\nu}\rho} + \gamma_{\mu\nu}\Bigl( \frac{3}{2}K^{\rho\sigma}K_{\rho\sigma} - \frac{1}{2}K^{2} \Bigr) + f \gamma_{\mu\nu} - 2\chi a_{\mu}a_{\nu} = 8\pi \gamma_{\mu}^{\ph{\mu}\rho} \gamma_{\nu}^{\ph{\nu}\sigma} T_{\rho\sigma}\; .
\end{align}
Finally, making use of the correspondence [compare also with \eqref{fluid}] 
\begin{subequations}\label{corresp}
\begin{align}
	\varepsilon &= N^{2} T^{00}  \;,\\
	J^{i} &= N \bigl( T^{0i}+N^{i}T^{00} \bigr)  \;, \\
	\mathcal{T}^{ij} &= T^{ij}-N^{i}N^{j}T^{00}  \;,
\end{align}
\end{subequations}
one can check that, when using the adapted 3+1 coordinates, the equations \eqref{ham1}, \eqref{constr1} and \eqref{evol1} are the same as the 3+1 equations \eqref{ham}--\eqref{evol} obtained by direct variation of the 3+1 Lagrangian \eqref{L3D}.

\section{Stress-energy tensor of the equivalent matter field}\label{appB}

As we discussed in Section \ref{secII}, we can consider the $\tau$-field, in the $4d$ formulation, as an additional matter field. Its stress-energy-momentum tensor is then 
\begin{equation}\label{TlambdaApp}
	T^{\mu\nu}_{\tau} = - \frac{1}{8\pi}\Bigl[ f(a) g^{\mu\nu} + 2n^{\mu}n^{\nu}\nabla_{\rho}\bigl(\chi(a) a^{\rho}\bigr) - 2\chi(a) a^{\mu}a^{\nu}\Bigr] \; .
\end{equation}
If we choose $n^{\mu}$ as the natural 4-velocity associated with this matter fluid, we can perform the canonical decomposition of the energy-momentum tensor according to 
\begin{equation}
	T^{\mu\nu}_{\tau} = \bigl(\varepsilon_{\tau}+P_{\tau}\bigr)\,n^{\mu}n^{\nu} + P_{\tau}g^{\mu\nu} + 2Q_{\tau}^{(\mu}n^{\nu)} + \Sigma_{\tau}^{\mu\nu}\;,
\end{equation}
where $\varepsilon_{\tau}$ and $P_{\tau}$ denote the energy density and pressure of the fluid, where $Q_{\tau}^\mu$ is the heat flow orthogonal to the 4-velocity, \textit{i.e.} $n_{\mu}Q_{\tau}^{\mu} = 0$, and where $\Sigma_{\tau}^{\mu\nu}$ is the symmetric anisotropic stress-tensor which is transverse to the velocity and traceless, \textit{i.e.} $n_{\nu}\Sigma_{\tau}^{\mu\nu} = 0$ and $g_{\mu\nu}\Sigma_{\tau}^{\mu\nu} = 0$. We then obtain 
\begin{subequations}
\begin{align}
	\varepsilon_{\tau} &= \frac{1}{8\pi} \Bigl[ f -2\nabla_{\mu}\bigl(\chi a^{\mu}\bigr) \Bigr]\;,  \\
	P_{\tau} &= \frac{1}{8\pi} \Bigl[ -f + \frac{2}{3}a^2\chi \Bigr]\;, \\
	Q_{\tau}^{\mu} &= 0 \;,\\
	\Sigma_{\tau}^{\mu\nu} &= \frac{\chi(a)}{4\pi} \Bigl[ a^{\mu}a^{\nu} - \frac{1}{3}\gamma^{\mu\nu}a^2 \Bigr]\;.
\end{align}
\end{subequations}

In terms of those definitions, the 3+1 version of the conservation equation $\nabla_\nu T_{\tau}^{\mu\nu}=0$ reduces to a scalar equation, taking the form of the ``continuity'' equation
\begin{equation}\label{contlambda}
	D_t \varepsilon_{\tau} + N\bigl(\varepsilon_{\tau}+P_{\tau}\bigr)K + N\Sigma_{\tau}^{ij}K_{ij} = 0 \; .
\end{equation}
As we mentioned in Sections \ref{secII} and \ref{secIII}, the conservation equation $\nabla_\nu T_{\tau}^{\mu\nu}=0$ has exactly the same content as the Khronon field equation \eqref{Teq}, and is equivalent, \textit{via} the modified Einstein equations \eqref{EE}, to the conservation of matter $\nabla_\nu T^{\mu\nu}=0$. Thus, this equivalent fluid has the property that its Euler equations reduce to its continuity equation, \textit{i.e.} the four components of $\nabla_\nu T_{\tau}^{\mu\nu}=0$ reduce to only one equation (as expected for a scalar field).

\bibliography{/home/blanchet/Articles/BM11/ListeRef.bib}

\end{document}